\documentclass[prb,preprint,amsmath,amssymb,superscriptaddress,notitlepage]{revtex4-2}
\usepackage{epsfig}
\usepackage{graphicx}
\usepackage{color}
\usepackage{soul}
\usepackage{hyperref}
\usepackage{bm}
\usepackage{fixme}

\usepackage{color,soul}
\usepackage{dcolumn} 
\usepackage{amsfonts}
\usepackage{natbib}
\usepackage{upgreek}
\usepackage{amssymb}
\usepackage{amsmath}
\usepackage{physics}
\usepackage{graphicx}
\usepackage{lipsum}
\usepackage{float}
\usepackage[dvipsnames]{xcolor}
\usepackage{url}

\usepackage[colorinlistoftodos,textwidth=1.5cm]{todonotes} 

\usepackage{xr}
\usepackage[utf8]{inputenc}

\usepackage{xargs}
\newcommandx*{\mypartial}[3] [3]{
 \frac{\partial^{#3} #1} {\partial #2 ^{#3}  }
}

\pdfoutput=1

\usepackage[normalem]{ulem}
\hypersetup{
    colorlinks=true,
    urlcolor=blue,
    linkcolor=blue,
    citecolor=blue}
\usepackage{upgreek}

\newcommand{\otherlabel}[2]{\protected@edef\@currentlabel{#2}\label{#1}}

\newcommand{\secref}[1]{Supp. Inf.~\ref{#1}}

\begin{document}

\title{Chaotic Proliferation of Relativistic Domain Walls for Reservoir Computing}

\author{J. A. V\'elez}
\email{jvelez008@ikasle.ehu.eus}
\affiliation{Donostia International Physics Center, 20018 San Sebasti\'an, Spain}
\affiliation{Polymers and Advanced Materials Department: Physics, Chemistry, and Technology, University of the Basque Country, UPV/EHU, 20018 San Sebasti\'an, Spain}

\author{M.-K. Lee}
\affiliation{Department of Applied Physics, Waseda University, Okubo, Shinjuku-ku, Tokyo 169-8555, Japan.}

\author{G. Tatara}
\affiliation{RIKEN Center for Emergent Matter Science (CEMS) and RIKEN Cluster for Pioneering Research (CPR), 2-1 Hirosawa, Wako, Saitama, 351-0198 Japan}

\author{P.-I. Gavriloaea}
\affiliation{Instituto de Ciencia de Materiales de Madrid, CSIC, Cantoblanco, 28049 Madrid, Spain\looseness=-1}

\author{J. Ross}
\affiliation{School of Physics, Engineering and Technology, University of York, York YO10 5DD, UK.}

\author{D. Laroze}
\affiliation{Instituto de Alta Investigación, CEDENNA, Universidad de Tarapacá, Casilla 7D, Arica, Chile.\looseness=-1}
 
\author{U. Atxitia}
\affiliation{Instituto de Ciencia de Materiales de Madrid, CSIC, Cantoblanco, 28049 Madrid, Spain\looseness=-1}

\author{R. F. L. Evans}
\affiliation{School of Physics, Engineering and Technology, University of York, York YO10 5DD, UK.}

\author{R. W. Chantrell}
\affiliation{School of Physics, Engineering and Technology, University of York, York YO10 5DD, UK.}
\author{M. Mochizuki}
\affiliation{Department of Applied Physics, Waseda University, Okubo, Shinjuku-ku, Tokyo 169-8555, Japan.}

\author{R. M. Otxoa}
\email{ro274@cam.ac.uk}
\affiliation{Hitachi Cambridge Laboratory, J. J. Thomson Avenue, Cambridge CB3 0HE, United Kingdom\looseness=-1}
\affiliation{Donostia International Physics Center, 20018 San Sebasti\'an, Spain}

\date{\today}

\begin{abstract}
\bf{
Magnetic domain walls in antiferromagnets have been proposed as key components for faster conventional information processing, thanks to their enhanced stability and ultrafast propagation. However, how non-conventional computing methods like reservoir computing might take advantage of these properties remains an open question. In this work, we show how complex domain wall patterns can form through the proliferation of multiple domain walls from the energy stored in a single seed domain wall driven to move at a high speed close to the relativistic limit. We demonstrate that the resulting magnetic texture, consisting of up to hundreds of domain walls with an overall conserved topological charge as the initial seed domain wall, can possess chaotic spatiotemporal dynamics depending on the strength of staggered spin-orbit field induced via applied current. These findings allow us to design a multiple-domain-wall reservoir with high short-term memory and nonlinearity with respect to spin-orbit field inputs, that is suitable for ultrafast, energy-efficient, non-conventional reservoir computing.
}
\end{abstract}

\bigskip

\maketitle


\section{Introduction}

Antiferromagnetic materials (AFMs) have garnered significant attention due to their unique properties and potential applications in spintronic devices. Unlike ferromagnetic materials, AFMs exhibit no net magnetization, resulting in minimal parasitic fields and superior stability against external magnetic disturbances~\cite{Baltz2018,MacDonald2011,Jungwirth2018,Jungfleisch2018,Fukami2020}. This inherent advantage makes AFMs highly suitable for applications where magnetic noise needs to be minimized.

Since the nineties, the study of the mobility of domain walls (DWs), vortices and skyrmions in ferromagnets has been the flagship in spintronics because of its potential in data storage~\cite{parkin2008magnetic} and logic devices~\cite{allwood2005magnetic}. However, the principal difficulty in leveraging these technological proposals arises from the intrinsic instability of these magnetic topological defects when they exceed a certain threshold velocity. In the case of DWs in ferromagnets, this undesired dynamical regime is known as the Walker breakdown~\cite{schryer1974motion} and features the dynamics that combine translational and oscillatory motions~\cite{mougin2007domain}. Antiferromagnetic (AF) textures are not exempt from an upper limit of speed associated with the maximal group velocity of magnons, $v_{\text{g}}$~\cite{shiino2016antiferromagnetic,gomonay2016high,otxoa2020giant}. This behavior stems from the Lorentz-invariant Lagrangian in AFMs that allows for intriguing relativistic kinematics for AF DWs, with the role of photons in conventional spacetime being replaced by magnons in magnetic systems~\cite{kim2014propulsion,rama2020steady}. However, the robustness of the DW structure is compromised when approaching the limiting speed. In particular, complex dynamical behavior, with clear Walker breakdown signatures similar to those found in ferromagnets, has also been observed in theoretical studies of layered AFMs~\cite{Otxoa2020b,lee2023predicted,Cheng2016}. 
\newline\indent
Moreover, it has recently been theoretically proposed that a DW moving at relativistic speeds can produce a pair of magnons~\cite{tatara2020magnon}, in an analogous fashion with the vacuum polarization by a strong electric field in quantum electrodynamics known as the Schwinger effect~\cite{schwinger1951gauge}. In general, the energy required to observe this phenomenon is typically achievable only with large-scale experimental facilities like Hadron colliders and ultrahigh-intensity laser light sources. However, in the field of condensed matter, indirect evidence of such an effect has been reported in graphene, where the low system dimensionality allows for an increased rate of production of electron-hole pairs~\cite{allor2008schwinger,schmitt2023mesoscopic}, in analogy with the electron-positron pair production predicted in Schwinger effect. In an AFM, the role played by the electric field is actually taken on by the moving DW, where the energy shift due to the Doppler effect in the laboratory frame matches the energy gap for the magnon creation, leading to a spontaneous emission. Regardless of the type of the driving mechanism — either spin-orbit torque (SOT)~\cite{Otxoa2020b}, spin-transfer torque~\cite{lee2023predicted} or the recently proposed laser-optical torque~\cite{gavriloaea2024efficient} — nucleation of DWs with opposite topological charges has been systematically predicted theoretically. This magnon pair emission not only enhances our fundamental understanding of relativistic spin dynamics but also has potential practical implications for developing novel spintronics devices. However, a deeper understanding of the nucleation process remains an open question. This issue is particularly important because AF DWs moving at relativistic speed can be compressed to a few nanometers and transport energy in a fast manner~\cite{otxoa2021topologically,otxoa2023tailoring}, making them promising for potential device applications. Most studies to date do not explore the influence of the time-dependent driving mechanism, the problem whether the proliferation of DWs is a chaotic process both in time and space, or ultimately the potential for generating DWs patterns to perform computational tasks inspired by bio-systems.
\newline\indent
In this work, supported by atomistic spin dynamics simulations, we theoretically show that from a seed of relativistically moving AF DW, a pattern of multiple pairs of AF DWs can be controllably generated under the application of time-varying electric current. The spatiotemporal dynamics of DWs driven by the staggered spin-orbit field induced by ac current in layered AFMs such as Mn$_2$Au or CuMnAs~\cite{wadley2016electrical,vzelezny2014relativistic} where the two magnetic sublattices are related by inversion symmetry, is analysed in a model of one-dimensional ferromagnetic chains with mutual AF coupling. Specifically, the Landau-Lifshitz-Gilbert-Slonczewski (LLGS) equation is numerically solved for an initially stabilized single DW being driven by current-induced staggered spin-orbit field. Two distinct behaviors are observed depending on the strength of the spin-orbit field, with steady and periodic spin-wave emissions under weak fields and chaotic dynamics with multiple DW proliferations under strong fields. This chaotic proliferation of DWs can be understood as analogous to turbulence in fluids, where the proliferation of magnetic textures resembles the turbulent behavior propagating through space and time, driven by complex nonlinear interactions. 
\newline\indent
Based on this investigation, we propose and demonstrate the potential application of DW proliferation, interacting DWs, spin wave emission, and the high nonlinearity of the system to the reservoir computing (RC)~{\cite{Magnetic01,Kanao2019,STO02,STO03,Furuta2018,Nakane2018,Yamaguchi2020,Bourianoff2018,Prychynenko2018,Pinna2020,Jiang2019,Lee2022,lee2023handwritten,Yokouchi,Msiska}, which is a newly established machine-learning framework that replaces hidden layers in a recurrent neural network by a single reservoir \cite{Tanaka2019,Nakajima2020,Nakajima2018}. The reservoir, now proposed as a texture of multiple AF DWs, plays a crucial role of nonlinearly transforming input data into an output function to solve linearly inseparable tasks. We show that an array of multiple AF DWs can serve as a physical reservoir with high capacities of short-term memory (STM) and nonlinearity demonstrated in the STM and parity-check (PC) tasks, respectively, leveraging the magnetization responses excited by the input spin-orbit fields. The dependence of the performances on strength and pulse width of spin-orbit fields, as well as the position of detectors are studied systematically. These metrics provide insights into the optimization of AF-DW-based reservoirs. By combining theoretical insights from dynamical AF DW studies with practical applications in RC, this paper aims to advance the field of spintronics and open new pathways for developing efficient, energy-conserving, and high-speed computational devices.

\section{Results and Discussion}

\subsection{Dynamics of Domain Walls}

We consider two one-dimensional ferromagnetic tracks with mutual AF exchange coupling. Within each track, the initial classical magnetization configuration, $\mathbf{M}(x,t)$, at position $x$ along the track and time $t$, is taken as a single steady DW, with their magnetizations being opposite in direction in the two tracks. After applying an ac current in $x$ direction to induce the ac staggered spin-orbit field (with amplitude proportional to current~\cite{wadley2016electrical} and frequency assumed to be the same as the ac current) that points in opposite directions in the two tracks at each time (see Methods), the dynamics of the $x$ component of the DW texture in the upper track, $M_x$, along the track as a function of time is illustrated in Fig. \ref{fig_1}, highlighting the differences between periodic and chaotic states in an AF system under a time-oscillating staggered spin-orbit field. In this work, the current-induced spin-transfer torque is ignored, as its effect on DW motion is prevailed by the staggered spin-orbit field, as shown recently in~\cite{lee2023predicted}. Panel \ref{fig_1}a shows a periodic state, where the DW oscillates regularly under a spin-orbit field amplitude of $H_{\text{SO}} = 745 \ \text{Oe}$ with a period of $25 \ \text{ps}$. The inset at the bottom of panel \ref{fig_1}a illustrates the initial condition, depicting a 180° DW configuration and its associated topological charge $Q$ defined by
\begin{equation}
Q = -\frac{1}{2\pi} \int_{-\infty}^{\infty} \nabla \phi(x,t) \, dx,
\end{equation}
where $ \phi(x,t) $ represents the angle of the magnetization in the plane of each track. This topological number counts how many times the magnetization wraps around the unit circle along the DW \cite{Braun2012}, which is conserved during the time evolution of the system, ensuring that the DW structure is topologically protected and maintains its stability against external perturbations. 

The upper inset in the grey area of Fig.~\ref{fig_1}a reveals a magnified view of the DW oscillations, showing the emission of spin waves. These periodic oscillations lead to spin wave emissions due to Lorentz contraction, which transport energy and angular momentum across the system. According to Shiino et al.~\cite{Shiino2016}, as the DW velocity approaches the maximal group velocity of spin waves, $v_{\text{g}}$, Lorentz contraction induces spin wave emission in the terahertz frequency range. The relativistic contraction of DW width $\Delta$ is described by
\begin{equation}
\Delta(v) = \Delta_0 \sqrt{1 - \left(\frac{v}{v_{\text{g}}}\right)^2},
\end{equation}
where $\Delta_0 = 19.8 \ \text{nm}$ corresponds to the DW width in the static state. This equation illustrates that as the DW velocity increases, its width $\Delta(v)$ decreases. This contraction is evident in the periodic oscillations as shown in the upper inset in grey area of Figure~\ref{fig_1}a, where the width of the red-colored region along $y$-axis is larger near the motion-reversal instants at which DW velocity is smaller. The spin wave emission is also clearly observed in this simulation result. This approach highlights the connection between topology and the dynamics of DWs in AF systems, similar to what is observed in ferromagnetic systems.

Figure~\ref{fig_1}b shows a chaotic state induced by a spin-orbit field amplitude of $H_{\text{SO}} = 780 \, \text{Oe}$ with the same current oscillation period. In this regime, a complex behavior is observed where the initial DW undergoes nucleation and proliferation of new DWs. The dominant energy of the DW texture relevant to its width comes from ferromagnetic exchange ($\propto J_{\rm F}$) plus easy-axis anisotropy ($\propto K_{2\perp}$) energies along the track, and can be written as $E=(\gamma+1/\gamma)E_0$ with $E_0\propto\sqrt{J_{\rm F}K_{2\perp}}\propto J_{\rm F}/\Delta_0 \propto K_{2\perp}\Delta_0$  and $\gamma=1/\sqrt{1-v^2/v^2_{\rm g}}$ is the Lorentz factor. At initial static state, $\gamma=1$ and the energy is $2E_0$. When the DW starts to move with Lorentz width contraction, $\gamma$ increases from 1 and the energy increases from $2E_0$. Together with the energies created by the emitted spin waves, the accumulated energy increment until a critical point results in the DW breaking and subsequent nucleation of new DWs. This process reflects the dynamical instability in the system, where accumulated energy is released to create new domain structures \cite{Otxoa2020}. 

This phenomenon is distinct from spin wave emission since it involves the physical reconfiguration of DWs. Although spin waves do not directly cause nucleation, they influence this process by modulating local energy distribution along the track. Spin waves can interfere constructively or destructively, creating regions of high or low energy that affect DW stability and rupture points. This effect is evident in the figures, particularly in Fig~\ref{fig_1}b, where the magnified view in the lower left corner highlights the spin wave emissions that precede the nucleation of new DWs. Frequent DW collisions occur in the chaotic state, reconfiguring DWs, altering their trajectories and velocities, and potentially leading to new DW nucleation if collision energy is sufficient to overcome local energy barriers. The figures show areas where contour lines meet and mix, indicating DW collisions and reconfigurations. The energy distribution due to spin waves is crucial in DW dynamics, as spin waves act as an additional energy dissipation channel, allowing DWs to release accumulated energy. However, when this dissipation is insufficient, the remaining energy can lead to DW rupture and nucleation. The balance between spin wave emission and DW nucleation defines the complexity of the chaotic state observed. 

In summary, Fig.~\ref{fig_1} provides a comprehensive view of the transition between periodic and chaotic DW behavior in an antiferromagnetic system under an applied spin-orbit field. The distinction between spin wave emission and nucleation processes is clearly highlighted, offering a deep understanding of the dynamic mechanisms involved. Periodic oscillations result in spin wave emission that dissipates energy, whereas in the chaotic state, energy accumulation leads to DW rupture and nucleation. DW collisions and the influence of spin waves on energy distribution add another layer of complexity to the system's dynamics.

In order to capture the details of spatio-temporal dynamics of the system of DWs combined with spin waves, we use fast Fourier transform (FFT) to analyse the magnetizations as follows. Figures~\ref{fig_2}a and \ref{fig_2}b are enlarged views of $M_x$ corresponding to the periodic state in Fig.~\ref{fig_1}a and chaotic state in Fig.~\ref{fig_1}b, respectively. In Fig.~\ref{fig_2}a, stable dynamics are observed as the spin wave emission occurs in a repetitive manner in time with a fixed period. The FFT of time evolution of $M_x$ at each position into frequency space is shown in Fig.~\ref{fig_2}b, displaying distinct peaks characteristic of periodic motion. Specific oscillation nodes are observed, indicating both spatial and temporal regularity in the system. To confirm this expectation, the time evolution of $M_x$ averaged along the entire track is shown in Fig.~\ref{fig_2}c, and regular oscillations synchronized with the applied field is observed. The FFT of the spatially averaged $M_x$ presented in Fig.~\ref{fig_2}d corroborates the periodic magnetisation oscillations in time with a pattern of regular peaks in frequency domain highlighting the deterministic behavior of the DW dynamics.

On the other hand, Fig.~\ref{fig_2}e depicts the dynamics of $M_x$ in a chaotic state. The dynamics reveal the proliferation and annihilation of DWs and their complex interactions, indicating chaotic behavior in both space and time. The FFT of $M_x$ is presented in Fig.~\ref{fig_2}f, revealing peaks smeared in a broad and continuous distribution of frequencies, which is typical of a chaotic state that reflects the irregular and non-repetitive nature of magnetization dynamics. This complexity in frequency domain indicates the absence of a dominant pattern and shows a superposition of multiple oscillation modes that are interacting nonlinearly. This chaotic dynamics is highly sensitive to initial conditions and external perturbations. The presence of multiple peaks in the FFT spectrum suggests the existence of various temporal and spatial scales in the DW dynamics. The spatial average of $M_x$ being irregular in time as shown in Fig.~\ref{fig_2}g again confirms this behavior. The FFT of this average value,  presented in Fig.~\ref{fig_2}h, reveals a broad and complex frequency spectrum, indicating the nonlinear and chaotic dynamics of the system.

Figure~\ref{fig_3} illustrates the phase transitions of the dynamics of our AF DWs represented by a bifurcation diagram. To survey the dependence of magnetization dynamics on the excitation spin-orbit field, we calculate two primary dynamic indicators as the growth rate of new DWs, $\xi$, and the complexity, $C$, which allow us to characterize the transition between periodic and chaotic states as a function of various field amplitudes $H_{\text{SO}}$ from 650 to 850 Oe in an interval of 5 Oe, and periods $T = {15, 20, 25, 30} \ \text{ps}$. In Fig.~\ref{fig_3}a, we illustrate the chaotic dynamics of the DWs and their proliferation process. An analogy can be drawn to the turbulence model in fluid mechanics, where magnetic textures proliferate in space and time in a manner similar to vortexes in a turbulent fluid. This detailed view enables us to clearly observe the complex interactions and reconfiguration of the DWs. In Fig.~\ref{fig_3}b, we present the number of detected DWs as a function of time, represented by blue points. The pink points indicate the moving average of the number of DWs, calculated by averaging over a window of 100 time steps starting just when the DWs begin to nucleate that is, after a transient period during which the DW still oscillates without nucleation. This approach allows us to smooth out rapid fluctuations and highlight general trends in the proliferation of DWs. The green line shows a linear fit of these averaged points. The observed fluctuations highlight the nucleation and annihilation processes of DWs, emphasizing the nonlinear dynamics of the system.

Figure~\ref{fig_3}c shows the topological configuration of DWs at three instants labelled by I-III in Fig.~\ref{fig_3}a. Above each DW texture, their topological charge $Q$ is indicated, revealing that DWs are produced in pairs with opposite topological numbers, such that the net $Q$ is conserved in time as shown in the right panel of Fig.~\ref{fig_3}c. Hence, despite the chaotic proliferation and complex dynamics, the system remains in the same topological class at all times. In  Fig.~\ref{fig_3}d, we show the bifurcation diagram with the complexity $C$ indicated by blue and red spheres respectively for the chaotic and periodic states, and DW growth, as functions of the amplitude ($H_{\rm SO}$) and period ($T$) of the spin-orbit field. The complexity is a measure that combines entropy and disequilibrium to quantify the degree of order and disorder in the system (See \secref{sec:supplementary_material} for details), which is associated with the color of the spheres in the graph (blue if $C>0$, indicating chaos, and red if $C=0$, indicating periodicity)~\cite{lopez1995statistical}. On the other hand, a linear trend is observed in the growth rate $\xi$, suggesting a proportionality to the field amplitude $H_{\text{SO}}$ given by $\xi = a_i H_{\text{SO}} + b_{i}$. The slopes $a_i$ represent the sensitivity of the nucleation rate to changes in $H_{\text{SO}}$ for each specific excitation period, with values $a_i = \{1.0538, 0.786, 0.678, 0.556\} \ \text{Oe}^{-1}$ for $T = \{15, 20, 25, 30\} \ \text{ps}$, respectively. The phase transition observed in the figure is due to the nucleation process; when $\xi = 0$, no nucleation occurs, indicating a periodic state. Thus, a value of $\xi > 0$ reflects a nucleation process, implying chaos. However, we note that certain blue spots are observed at $\xi = 0$, reflecting that the system may be in a chaotic state even when the number of DWs is fixed as the initial state without proliferation of new DWs.

Altogether, Figs.~\ref{fig_1} to \ref{fig_3} offer a comprehensive view of how the ac staggered spin-orbit field influences both the nucleation rate and complexity of the system to induce the phase transition of the DW configuration and dynamics between periodic and chaotic phases. The nucleation process we observe is neither linear in time nor predictable, as it depends on local fluctuations of energy. The instability suffered by the relativistically moving DW disturbs its regular motion and texture, and eventually producing spin waves and new DWs to release energy, which interact with each other nonlinearly. This creates an environment where small variations can generate significant changes in the system dynamics and results in the chaotic behavior. In summary, the interplay of relativistic contraction, energy accumulation, and nonlinear interactions between DWs and spin waves induces chaotic dynamics during the nucleation process. This is reflected in the positive complexity observed in chaotic states, even in the absence of new DW proliferations. In these cases, chaos manifests through irregular and sensitive movements of the existing DWs, induced by nonlinear interactions and energetic fluctuations, without necessarily involving additional nucleations.

\subsection{Reservoir Computing}

After investigating the chaotic DW proliferation driven by staggered spin-orbit field in AFMs, in this section we propose the potential of a proliferated multiple-DW configuration as a viable system for spintronics reservoir computing (RC) by demonstrating the possession of short-term memory and nonlinearity inherent in the magnetization responses to external spin-orbit field inputs in this system. We consider two one-dimensional ferromagnetic layers with mutual AF exchange coupling, as illustrated in Fig.~\ref{fig_4}a. Each layer has a length of 6000 sites and contains seven equally-spaced DWs with centers located at site indices $750n$ with $n=1,2,...,7$ [see Fig.~\ref{fig_4}b]. Each neighboring DWs have opposite winding numbers, mimicking the texture generated by the chaotic proliferation from a single seed DW that preserves the topological number. We evaluate the performance by this system on two benchmark tasks of RC as the short-term memory (STM) and parity-check (PC) tasks~\cite{Kanao2019,Furuta2018,Lee2022}.

For both tasks, the input data $s_{\rm in}(T_i)$ consists of a sequence of random digits, either 1 or 0, at integer time steps $T_i$. The goal of RC by using this DW system is to predict specific transformations of the inputs by an output function, taken as a linear combination of the magnetization responses measured on local areas in the DW array. The correspondence between input digits and physical excitations in the DW system is designed as follows. When $s_{\rm in}(T_i)=1$, we simultaneously apply three local staggered spin-orbit field pulses in $\pm y$ direction for the upper and lower DW array, respectively. These pulses are applied to three input areas located on site indices from 1001 to 1500, 2500 to 3000, and 5001 to 5500, as indicated by green areas in Fig.~\ref{fig_4}b. When $s_{\rm in}(T_i)=0$, we reverse the directions of the spin-orbit field pulses for both layers and apply them on the same local areas with the same pulse width. The surveyed pulse width of spin-orbit field ranges from 0.1 to 0.5 ps, while its magnitude is varied from 10 to 60 mT in this study.

We placed eleven detectors on the DW array as shown in Fig.~\ref{fig_4}b. In the area of each detector, the averaged magnetization at a number of virtual-node temporal instants in each spin-orbit field pulse is recorded to define the reservoir state vector. A linear transformation of the measured magnetizations is defined as output function, with the coefficients being the weight vector components being optimized to minimize the mean square error between the target and output for a training set of random digits. The optimized weight vector is then taken to predict another testing set of digits by their magnetization responses (see details in Methods). The target functions for STM and PC tasks are respectively defined in Eq.~(\ref{STM}) and Eq.~(\ref{PC}) in Methods. From these definitions, the STM task investigates to what extent the input at a previous time $T_i-T_{\rm delay}$ can be reconstructed by the reservoir state at current time $T_i$, which is important for applications such as sentence prediction and speech recognition~\cite{Tanaka2019,Nakajima2020,Nakajima2018} that involve time-series data. Meanwhile, the PC task examines to what extent the reservoir can nonlinearly transform its components into a summation of past inputs modulo 2, which is essential for problems like pattern classification and hand-written digit recognition~\cite{lee2023handwritten,Yokouchi,Msiska} that require nonlinearity.


After the training procedure, the squared correlation between the testset targets and outputs for each detector is calculated. As shown in Supplementary Fig.~\ref{supp_fig_4}, empirically we find when a detector is placed close to both the edge of any input areas, and either to one of the DW centers (e.g., detectors 3 and 6 in Fig.~\ref{fig_4}b) or near the system edge (e.g., detectors 11), a better squared correlation is obtained for both tasks.
This indicates the need of both a large spatial gradient of $H_{\rm SO}$ and either a large gradient of $M_y$ or strong spin wave reflection near the system edge, to excite significant magnetization responses that possess memory and nonlinearity relative to the input.
To quantify the performance, we calculate the capacities $C_{\rm STM(PC)}$ for STM (PC) task~\cite{Kanao2019,Furuta2018,Lee2022} for the better detectors 3, 6, and 11, defined as the sum of the squared correlations from $T_{\rm delay}=0$ to $30$. A larger capacity indicates a better performance by the reservoir.
We achieve the highest $C_{\rm STM}$ and $C_{\rm PC}$ values of approximately 12.8 and 2.7, respectively, comparable with other spintronics reservoirs using a similar number of virtual nodes~\cite{Furuta2018,Kanao2019,Yamaguchi2020,Lee2022}. This finding clearly demonstrates the potential of a multiple-DW array proliferated by spin-orbit field in AFMs for applications in physical reservoir computing. We have tested the reproducibility by performing sequential runs of the tasks, and compared the performances between the proposed multiple-DW array and a pure AF state without DW textures and find the multiple-DW array shows much better results (see Methods for details).

To investigate potential input dependence of the performance, we compare the capacities carried out by different components of magnetizations under varying pulse widths and amplitudes of spin-orbit field, as illustrated in Fig.~\ref{fig_5}. In Fig.~\ref{fig_5}a, the pulse width is fixed at 0.2~ps for all curves. The results show that increasing the field amplitude from 10 to 60~mT does not  significantly change the capacities. For $H_{\rm SO}=10$ mT, the capacities carried out by $M_x$ and $M_y$ components were lower than those for higher SO-field amplitudes. 
Interestingly, the data reveals a contrasting behavior between the STM and PC capacities. Specifically, the $M_z$ component has roughly better results for STM capacities compared to $M_x$ and $M_y$. On the contrary, for PC capacities, $M_z$ is worse than the other components. This opposite behavior between STM and PC capacities is reminiscent of the empirical law of memory-nonlinearity trade-off~\cite{Dambre2012,Inubushi2017}, which is frequently observed in dynamical models or physical systems in RC, suggesting that the introduction of nonlinearity into reservoir dynamics tends to degrade memory capacity. 

In Fig.~\ref{fig_5}b, we study how the capacities for STM and PC tasks change with different pulse widths, ranging from 0.1 to 0.6~ps. For detectors 3 and 6, as the pulse width increases from 0.1 and 0.4~ps, the STM capacities tend to decrease, while the PC capacities increase. This opposite behavior is another indication of memory-nonlinearity trade-off, where an increase in one capacity typically leads to a decrease in the other. 
However, detector 11 exhibits a unique behavior. As the pulse width increases, both STM and PC capacities follow a similar trend, which violates the expected trade-off. This suggests that for detector 11, it may be possible to simultaneously enhance the short-term memory and nonlinearity by fine-tuning the input pulse width. This capability is particularly advantageous for machine learning tasks, since many realistic problems require both memory and nonlinearity. The common trade-off between memory and nonlinearity limits the learning potential of reservoirs, making our finding significant. 
This distinct behavior of detector 11 as compared to detectors 3 and 6 may be attributed to  edge effects of the system. These edge effects might enhance both memory and nonlinearity due to phenomena like spin wave reflection and interference occurring near the system edge. It is left for our future work to find concrete ways to overcome the memory-nonlinearity trade-off and to enhance capacities for both STM and PC tasks in this multiple-DW reservoir.

\section{Conclusions}

This study has elucidated the dynamical behavior of AF DWs under the influence of a time-varying staggered spin-orbit field, revealing the conditions of field amplitude and period that lead to either periodic or chaotic states. We have demonstrated that the transition to chaotic states are both chaotic in space and time, characterized by complex DW pattern formation and driven by the interplay of Lorentz contraction and spin wave emissions. These interactions lead to the proliferation of DWs from an initial seed DW, resulting in intricate patterns that reflect the chaotic dynamics in the system, quantified by the calculated complexity and DW proliferation rate. The findings also demonstrate that these chaotic DW dynamics offer significant potential for RC, as the AF DW array exhibits large short-term memory and nonlinearity capacities relative to the spin-orbit field input, essential for making them viable to tackle computational tasks. The ability of these AF-DW systems to achieve high performance in RC tasks underlines their promise as robust platforms for developing future spintronics-based computational technologies.

Furthermore, we have demonstrated the ability to reconfigure the DW reservoir to potentially solve tasks of varying difficulty level by adjusting the amplitude and period of the spin-orbit field.
This tunability allows the system to adapt its dynamic properties, thereby optimizing the performance of RC systems for specific tasks.
This work provides the basis of future fine-tuning of the spin-orbit field and excitation period to optimize the RC performance, and reveals an insight to potentially overcome the memory-nonlinearity trade-off, paving the way for the development of advanced, efficient, high-speed spintronic devices that leverage the dynamical behavior of AF DWs. This research significantly advances our understanding of AF-DW dynamics and their applications, laying the groundwork for the development of novel spintronic technologies. In summary, our work not only sheds light on the fundamental aspects of AF-DW dynamics but also paves the way for their practical application in cutting-edge spintronics reservoir computing.


\section{Methods}

\subsection{Theoretical Model}

The system under study is the layered collinear antiferromagnet, Mn\textsubscript{2}Au, characterized by its good conductivity, strong magnetocrystalline anisotropy, and a N\'eel temperature well above room temperature. These intrinsic properties make this material ideal for exploring the dynamics of antiferromagnetic domain walls due to its intrinsic properties.

The total energy of the system includes contributions from exchange interactions, magnetocrystalline anisotropy, and the spin-orbit field induced by an alternating current. The configurational energy $E$ can be expressed as:

\begin{align}
E &= - \sum_{\langle i,j \rangle} \mathcal{J}_{ij} \mathbf{M}_i \cdot \mathbf{M}_j - K_{2\perp} \sum_i (\mathbf{M}_i \cdot \hat{z})^2 \notag \\
  &\quad - K_{2\parallel} \sum_i (\mathbf{M}_i \cdot \hat{y})^2 - \frac{K_{4\perp}}{2} \sum_i (\mathbf{M}_i \cdot \hat{z})^4 \notag \\
  &\quad - \frac{K_{4\parallel}}{2} \sum_i \left[ (\mathbf{M}_i \cdot \mathbf{u}_1)^4 + (\mathbf{M}_i \cdot \mathbf{u}_2)^4 \right] \notag \\
  &\quad - \mu_0 \mu_s \sum_i \mathbf{M}_i \cdot \mathbf{H}^{i}_{\text{SO}}
\end{align}

where $ \mathbf{S}_i $ is the spin vector at site $i$, $\mathcal{J}_{ij}$ represents the exchange interaction between spins $i$ and $j$ with $\mathcal{J}_{1} = -(396 \, \text{K}) k_{\text{B}}$, $\mathcal{J}_{2} = -(532 \, \text{K} ) k_{\text{B}}$, $\mathcal{J}_{3} = (115 \, \text{K})  k_{\text{B}}$  ($k_{\text{B}}$ is the Boltzmann constant), $K_{2\perp}$ and $K_{2\parallel}$ are the second-order magnetocrystalline anisotropy constants in the perpendicular and parallel directions, respectively, with $K_{2\perp} = -1.303 \times 10^{-22} \, \text{J}$, $K_{2\parallel} = 7 \, K_{4 \parallel}$, $K_{4\perp}$ and $K_{4\parallel}$ are the fourth-order magnetocrystalline anisotropy constants in the perpendicular and parallel directions, respectively, with $K_{4\perp} = 2 \, K_{2 \parallel}$, $K_{4\parallel} = 1.855 \times 10^{-25} \, \text{J}$, and $\mathbf{H}^{i}_{\text{SO}} $ is the spin-orbit field at site $i$, induced by the alternating current. The unit vectors $\hat{u}_1$ and $\hat{u}_2$ represent the in-plane $xy$-based directions, with $\mathbf{u}_1 = [110]$ and $\mathbf{u}_2 = [1 \bar{1} 0]$. \\

The spin-orbit field ($\mathbf{H}^{i}_{\text{SO}} $) is induced by an alternating current flowing in the $ x $-direction. Due to the spin-orbit interaction, this current generates a spin-orbit field in the $ y $-direction. This field can be described as $ \mathbf{H}^{i}_{\text{SO}} = H_{\text{SO}} \sin(\omega t) \hat{y} $, where $ H_{\text{SO}} $ is the amplitude of the field, $ \omega $ is the angular frequency, and $ t $ is the time. This field introduces an additional interaction that affects the spin dynamics in the material. To describe the temporal evolution of the spins under the influence of the aforementioned fields, we use the Landau-Lifshitz-Gilbert (LLG) equation:

\begin{equation}
\frac{d\mathbf{M}_i}{dt} = -\gamma \mathbf{M}_i \times \mathbf{H}_i^\text{eff} - \gamma \alpha \mathbf{M}_i \times (\mathbf{M}_i \times \mathbf{H}_i^\text{eff})
\label{eq:LLG}
\end{equation} 

where $\gamma$ is the gyromagnetic ratio of the free electron (2.21$\times$10\textsuperscript{5} m/A s), $\alpha = 0.001$ is the Gilbert damping parameter, and $\mathbf{H}_i^\text{eff}$ is the effective field resulting from all the interaction energies. \\

The effective field $ \mathbf{H}_i^\text{eff} $ is obtained by differentiating the total energy $E$ with respect to $ \mathbf{S}_i$: \\

\begin{equation}
\mathbf{H}_i^\text{eff} = -\frac{1}{\mu_0 \mu_\text{s}} \frac{\delta E}{\delta \mathbf{M}_i}.
\label{eq:EffectiveField}
\end{equation}

This includes contributions from exchange interactions, anisotropy, and the spin-orbit field.

This theoretical model provides a solid foundation to explore the dynamics of domain walls in the antiferromagnetic system Mn\textsubscript{2}Au under the influence of an alternating current. By considering all energy contributions and using the Landau-Lifshitz-Gilbert equation, we can simulate and analyze how these factors influence the temporal evolution of the domain walls. This analysis  offers valuable insights for the development of reservoir computing applications and other advanced spintronic technologies.

\subsection{Reservoir Computing Method}

The commonly adopted target functions $y_{\rm target}$ for the STM and PC tasks are defined as~\cite{Kanao2019,Furuta2018,Lee2022}
\begin{align}
&y_{\rm target}^{\rm STM}(T_i,T_{\rm delay})=s_{\rm in}(T_i-T_{\rm delay}),
\label{STM}\\
&y_{\rm target}^{{\rm PC}}(T_i,T_{\rm delay})=[s_{\rm in}(T_i) + s_{\rm in}(T_i-1)\nonumber\\
&\hspace*{3cm} + \cdots +s_{\rm in}(T_i-T_{\rm delay})]~{\rm mod}~2,
\label{PC}
\end{align}
where $T_{\rm delay}$ is a dimensionless integer delay time, and $s_{\rm in}(T_i-T_{\rm delay})$ is the input digit (0 or 1) at a previous integer time of $T_i-T_{\rm delay}$.
We placed 11 detectors located at sites $501+500n$ with $n=0,1,...,10$ on the DW array as shown in Fig.~\ref{fig_4}(b). All detectors have a short length of 11 sites that corresponds to roughly 3.7~nm taking the lattice constant of Mn$_2$Au, which is nearly $\Delta_0/5$, in order to capture the possible variation of the performance carried out by magnetizations located at different positions inside or outside the DW region. For each detector, we measured their averaged magnetization at a number of $N_{\rm vn}$ virtual-node temporal instants in the interval from $T_ip$ to $(T_i+1)p$ with $p$ being the spin-orbit field pulse width, to define a reservoir state vector at time $T_ip$ as $\mathbf{R}^{(n)}_j(T_i)\equiv(\langle M^{(n)}_j(T_ip+p/N_{\rm vn})\rangle, \langle M^{(n)}_j(T_ip+2p/N_{\rm vn})\rangle, ..., \langle M^{(n)}_j((T_i+1)p)\rangle)$, with $j=x,y,z$ labeling the component of eleven magnetizations $\mathbf{M}^{(n)}$ inside the $n$th detector area, and $\langle ...\rangle$ denoting the average over the eleven sites within this detector. The scalar output function $y^{(n)}_{j,\rm out}$ is defined as a dot product between this $N_{\rm vn}$-dimensional reservoir vector and a weight vector with the same dimension, $\mathbf{W}^{(n)}_j$, then plus a constant bias $W^{(n)}_{j,0}$, namely $y^{(n)}_{j,\rm out}=\mathbf{W}^{(n)}_j\cdot \mathbf{R}^{(n)}_j+W^{(n)}_{j,0}$. The number of components in weight vector that are required for training is thus $N_{\rm vn}+1$, including one for the constant bias. Note that the weight vectors $\mathbf{W}^{(n)}_j(T_{\rm delay})$ differ among the delay times $T_{\rm delay}$,  the eleven detector positions, and the magnetization components $j=x,y,z$. They are required to be trained independently for each delay time,  each detector, and each component.

In simulation, the DW system is initially relaxed for a sufficiently long time until magnetizations are almost fixed in time. Then for each delay time $T_{\rm delay}=0,1,...$ separately, we stabilize the response of this system by first injecting 1000 random inputs of 1 or 0 via the corresponding sequence of SO-field pulses, then we take the following 100 random inputs for training, and their subsequent 50 inputs for testing. The training is done by minimizing the mean square error between the targets in Eq.~(\ref{STM}-\ref{PC}) and output function $y^{(n)}_{j,\rm out}$ using the pseudo-inversed matrix method~\cite{Fujii2017,Strang1993} for the 100 training dataset, after which, the optimal weight vector is used to form the output function of another distinct 50 testing inputs. The squared correlation between target and output~\cite{Kanao2019,Furuta2018,Lee2022} is then calculated as
\begin{eqnarray}
\text{Corr}^{(n,j)}_{\rm STM(PC)}(T_{\rm delay})=\frac{\text{Cov}^2[y^{\rm STM(PC)}_{\rm target}(T_i,T_{\rm delay}),y^{(n)}_{j,\rm out}(T_i)]}{\text{Var}[y^{\rm STM(PC)}_{\rm target}(T_i,T_{\rm delay})]\text{Var}[y^{(n)}_{j,\rm out}(T_i)]},
\end{eqnarray}
with
\begin{eqnarray}
&&\text{Cov}[A(T_i),B(T_i)]=\frac{1}{N}\sum_i (A(T_i)-\bar{A})(B(T_i)-\bar{B}),
\nonumber\\
&&\text{Var}[A(T_i)]=\frac{1}{N}\sum_i (A(T_i)-\bar{A})^2,
\end{eqnarray}
for generic functions $A$ and $B$, where Cov and Var denote the covariance and variance, respectively, $\bar{A}$ is the average of $A(T_i)$ over all $T_i$, and $N$ is the number of time steps $T_i$. The standard squared correlation $\text{Corr}$ takes a value within a range of [0,1], and a larger value indicates better fitting of the targets by outputs.

To quantify the performance of our DW-array reservoir, we calculate the capacity, which is defined as the summation of squared correlations from $T_{\rm delay}=0$ to $30$, excluding contributions from the delayed memory with peaks located at finite $T_{\rm delay}$ (see Supplementary Material), as shown in Fig.~\ref{fig_5} in the main text. A larger capacity indicates that a larger amount of memory or nonlinearity is stored in the reservoir state at current time $T_i$.

\section{Supplementary Material}
Detailed calculations, including the derivation of complexity and its application, are provided in the Supplementary Material, available online with this article.
\label{sec:supplementary_material}

\newpage

{\bf Acknowledgements}

JR and PG acknowledge funding from the European Union’s Horizon 2020 research and innovation programme under the Marie Skłodowska-Curie International Training Network COMRAD (grant agreement No 861300). M.M. and M.-K.L. thank the supports from Japan Society for the Promotion of Science KAKENHI (Grant No.~20H00337, No.~22H05114 and No.~24H02231), CREST, the Japan Science and Technology Agency (Grant No.~JPMJCR20T1), and a Waseda University Grant for Special Research Projects (Project No.~2024C-153). 
UA gratefully acknowledges support by grant PID2021-122980OB-C55 and the grant RYC-2020-030605-I funded by MCIN/AEI/10.13039/501100011033 and by "ERDF A way of making Europe" and "ESF Investing in your future".

{\bf Author contributions}

 R.M.O conceived the idea and designed the study. J.A.V. and R.M.O. carried out the atomistic spin dynamics simulations. M-K.L. and M.M. facilitated the understanding and implementation of the reservoir computing model. G.T.,  P.-I.G., J.R., R.E. and R.W.C. provided valuable comments and interpretation of the atomistic spin dynamics results.  J.A.V, D.L. and R.M.O developed the theoretical framework for the characterization of the chaotic dynamics. J.A.V., M-K.L. and R.M.O. wrote the initial draft and prepared the figures.  All authors reviewed the manuscript and provided comments and/or edited the final version of the draft.


\bibliography{bib2}
\bibliographystyle{unsrt}

\newpage

\noindent {\bf Figure Captions}
\\
\noindent {\bf Figure~\ref{fig_1}.}
{\textbf{Dynamic States of Domain Walls.}} {(a) Dynamics of $M_x$ along the upper track as a function of time for a periodic state, with the value represented in a color gradient (gray near $M_x=0$, pink near $M_x=1$, and cyan near $M_x=-1$). The upper panel shows the applied spin-orbit field $H_{\text{SO}} = 745$ Oe with a period of 25 ps. The lower inset illustrates the initial condition of the system, obtained by relaxing the domain wall, as well as the value of its topological charge ($Q = 1/2$) and the form of the AC current applied along the $x$ axis ($I_{Appl}$). The middle inset displays a magnified view of the oscillations in the DW, revealing the emission of spin waves due to the relativistic contraction of DWs. (b) Dynamics of the $S_x$ component along the track as a function of time for a chaotic state. The upper section shows the applied spin-orbit field $H_{\text{SO}} = 780$ Oe with a period of 25 ps. A nucleation process is observed when the DW reaches its maximal group velocity, causing the DW to contract following Lorentz transformations until a point where nucleation occurs, giving rise to new DWs while maintaining the same topological charge. The middle insets show magnified views of the DW proliferation process. The conservation of $Q$ ensures that, despite the creation and annihilation of multiple DWs, the total topological charge of the system remains constant, which is fundamental for integrity of the system.
}

\vspace*{1cm}
\noindent {\bf Figure~\ref{fig_2}.}
{\bf Soliton Proliferation.} {(a) Dynamics of $M_x$ along the upper track as a function of time for a periodic state as an enlarged view of Fig.~\ref{fig_1}a. 
(b) Fast Fourier Transform (FFT) of the temporal evolution of $M_x(x,t)$ at each spatial position from panel (a), transforming the time-space data into frequency-space. Periodic nodes are observed in both time and position. 
(c) Mean value of $M_x$ along the upper track as a function of time, showing the stability of the dynamics in the periodic state. 
(d) FFT of the mean value of $M_x$, indicating global periodic behavior. (e) Enlarged view of $M_x$ for the chaotic state in Fig.~\ref{fig_1}b.
(f) FFT of $M_x$ in panel (e), showing a broad spectrum of frequencies characteristic of chaotic dynamics. 
(g) Mean value of $M_x$ as a function of time for the chaotic state, highlighting the complexity of the dynamics. 
(h) FFT of the mean value of $M_x$, revealing the broad frequency spectrum typical of chaotic states.
}
\newpage
\vspace*{1cm}

\noindent {\bf Figure~\ref{fig_3}.}
{\bf Phase diagram.} {(a) Illustration of the chaotic dynamics of the DW and its proliferation process, using an analogy with the vortex model in fluid mechanics. This model facilitates the visualization and understanding of the complexity of DW behavior in the AF system. Labels I, II, and III indicate specific moments at which the topological configuration is schematically shown in panel (c). 
(b) Number of DWs detected as a function of time. The blue points represent the number of DWs detected at each time instant. The pink points show the moving average of the number of DWs, which smooths out fluctuations and allows for a better interpretation of the overall trend. The green line is a linear fit of the moving average points, indicating the growing trend of the number of DWs over time. 
(c) Topological configuration of the DWs at instances I, II, and III marked in panel (a). Each row shows the DWs at different time instants, highlighting the conservation of the topological charge $Q$. The numerically extracted time dependence of the topological charge is illustrated on the right side of the panel, showing that $Q$ is conserved over time despite the proliferation of new DWs. The color code represents the magnitude of $M_x$ with blue for $M_x = -1$, red for $M_x = 1/2$, and gray for $M_x = 0$.
(d) Bifurcation diagram using two dynamic indicators: complexity (with blue and red colors respectively representing chaotic and non-chaotic states) and the slope $\xi$, which is calculated from a linear fit of the moving average of the number of DWs as a function of time and represents the growth rate of the average number of proliferating DWs. This diagram is represented for four values of periods $T$ with the field amplitude $H_{\text{SO}}$ being increased from 650 Oe to 850 Oe in an interval of 5 Oe, allowing for observation of transitions between chaotic and non-chaotic states.
}

\vspace*{1cm}

\noindent {\bf Figure~\ref{fig_4}.}
{\bf Schematics RC.} {(a) Schematics of the bilayer AF DW system for RC. Black arrows indicate the local magnetization vectors and the green thick arrows are the local input SO-fields. (b) Initial configuration of the $x$-component of the DW magnetizations in the upper layer (blue curve). For the lower layer, the sign of $M_x$ is opposite. Green shaded zones are input areas and the eleven detectors are labeled by red boxes.
}

\newpage
\vspace*{1cm}

\noindent {\bf Figure~\ref{fig_5}.}
{\bf RC results. }{Comparison of the capacities as functions of (a) SO-field amplitude, and (b) SO-field pulse width for the three best detectors: 3, 6, and 11.}

\newpage

\begin{figure}[h!]
\hspace*{-0cm}
\includegraphics[width=1.\textwidth]{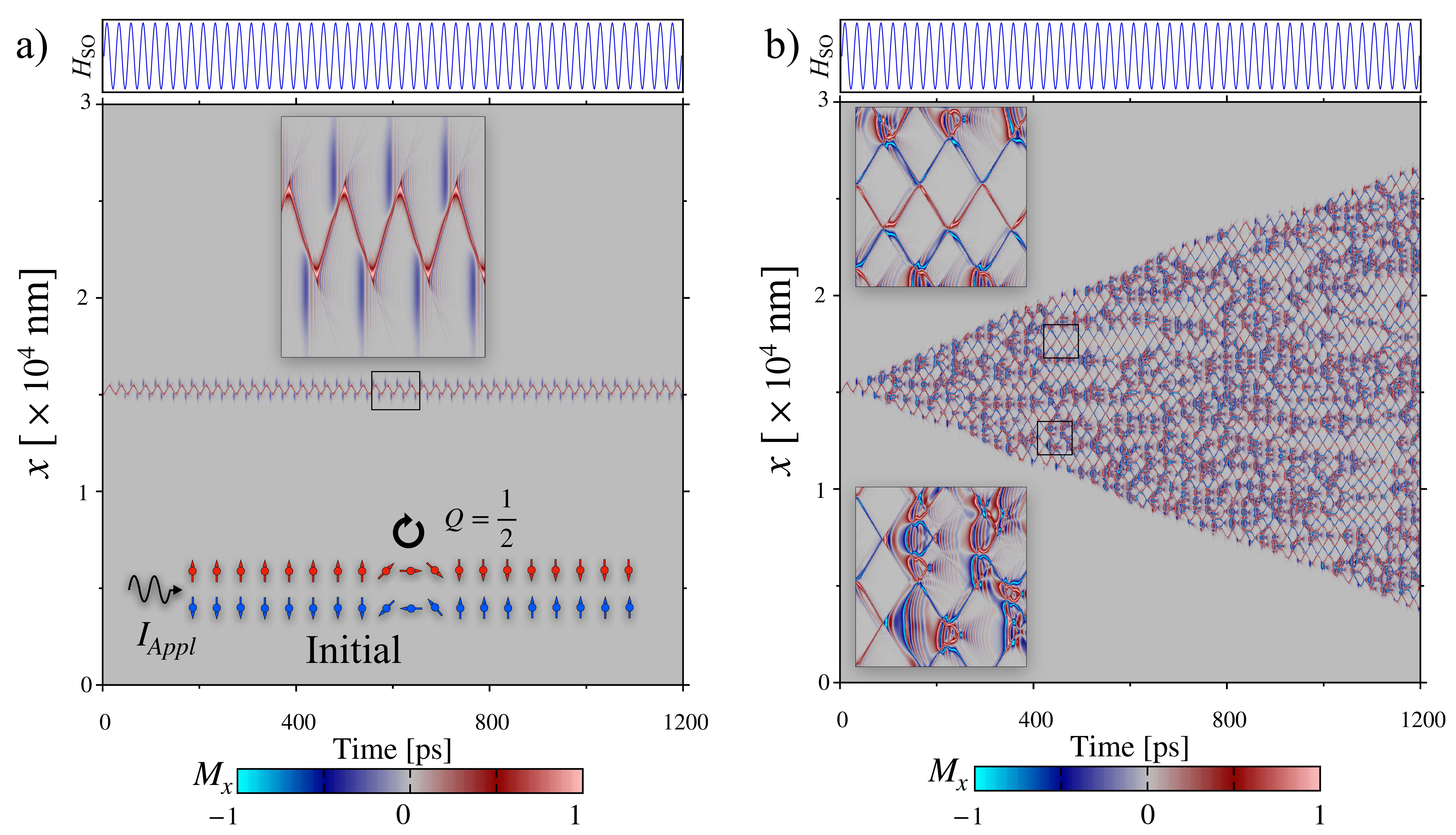}
\caption
{
}
\label{fig_1}
\end{figure}

\newpage

\begin{figure}[h!]
\hspace*{-0cm}\includegraphics[width=.99\textwidth]{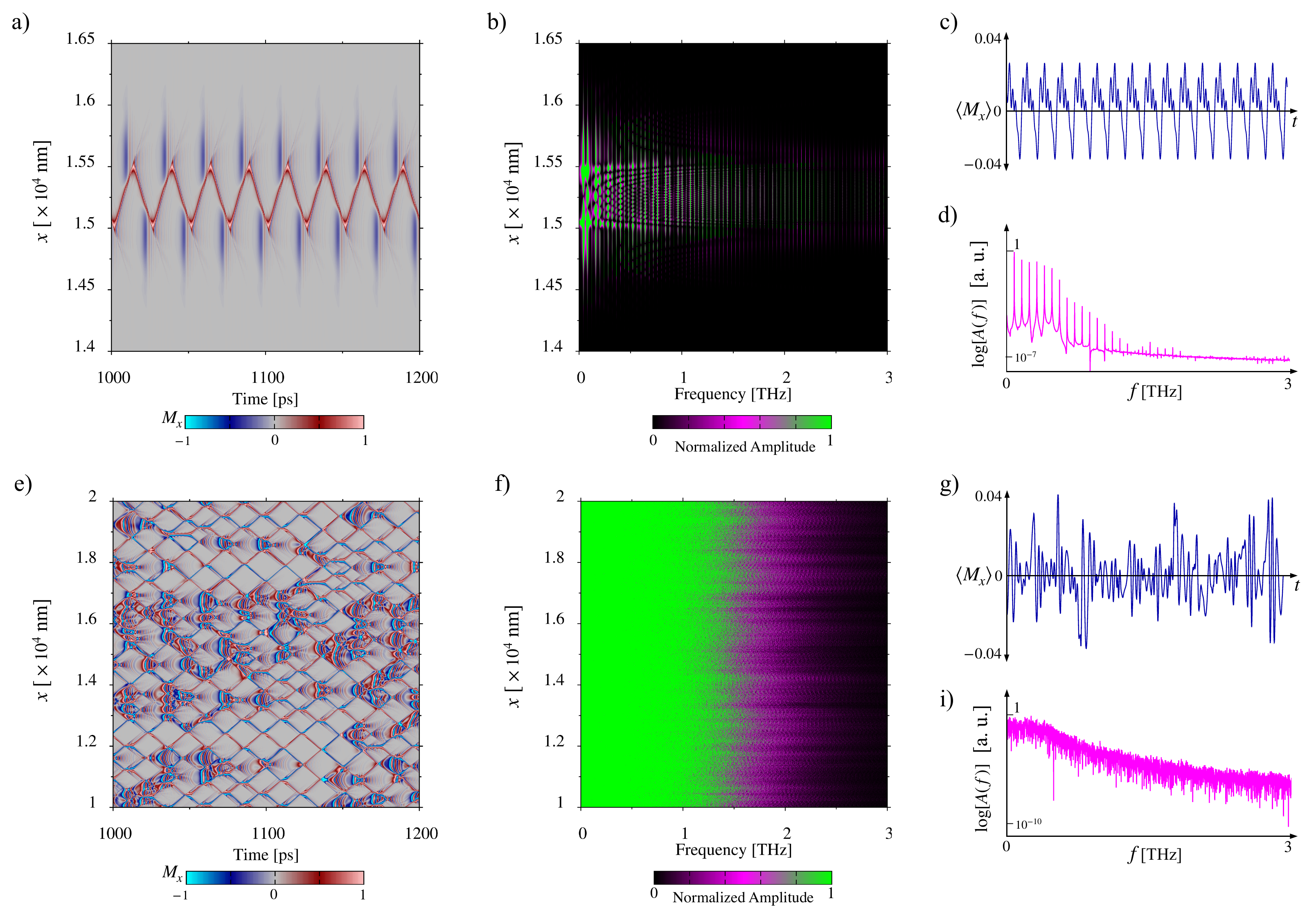}
\caption
{
}
\label{fig_2}
\end{figure}

\newpage

\begin{figure}[h!]
\hspace*{-0cm}\includegraphics[width=.99\textwidth]{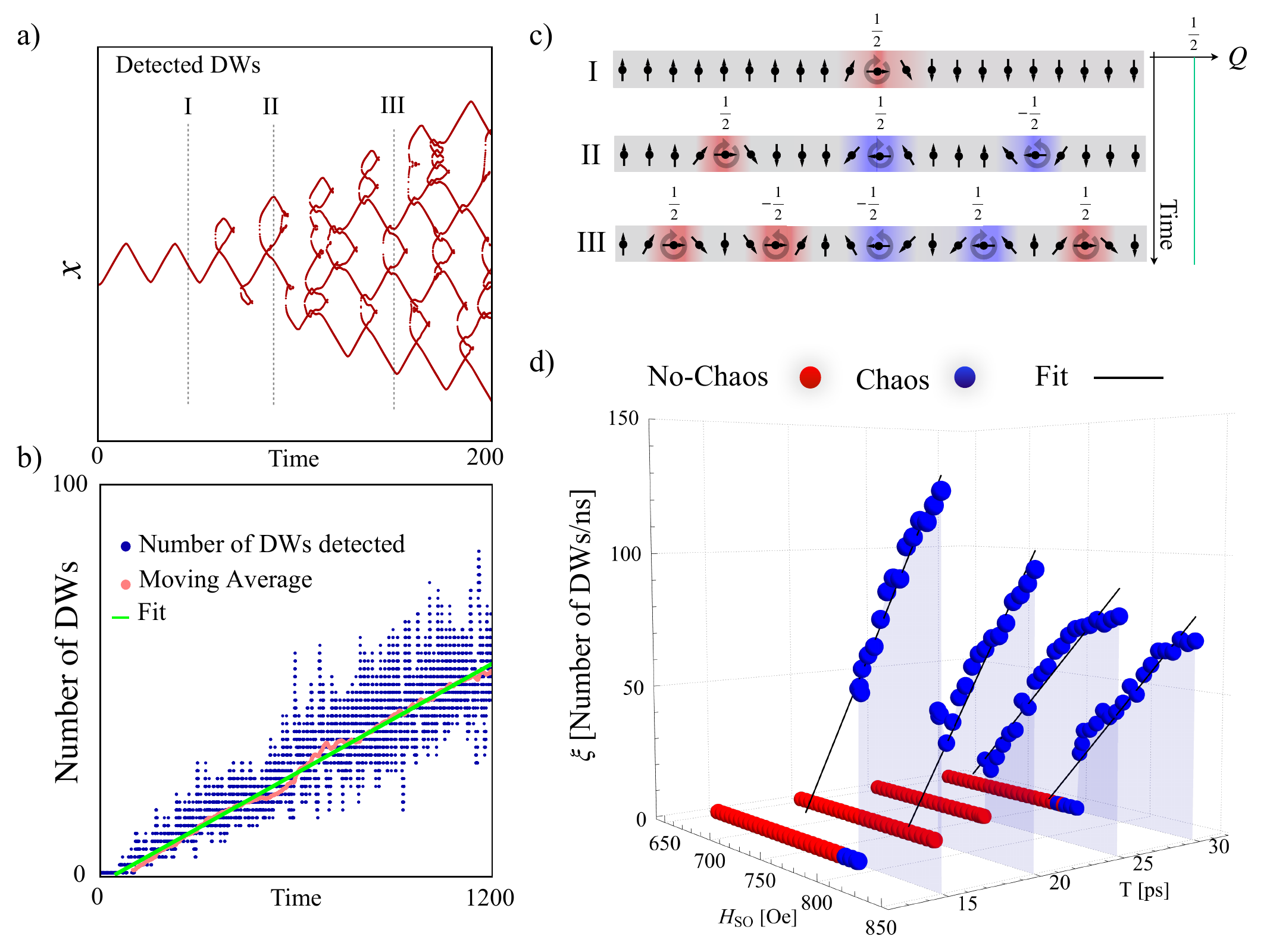}
\caption
{
}
\label{fig_3}
\end{figure}

\newpage

\begin{figure}[h!]
\hspace*{-0cm}\includegraphics[width=.99\textwidth]{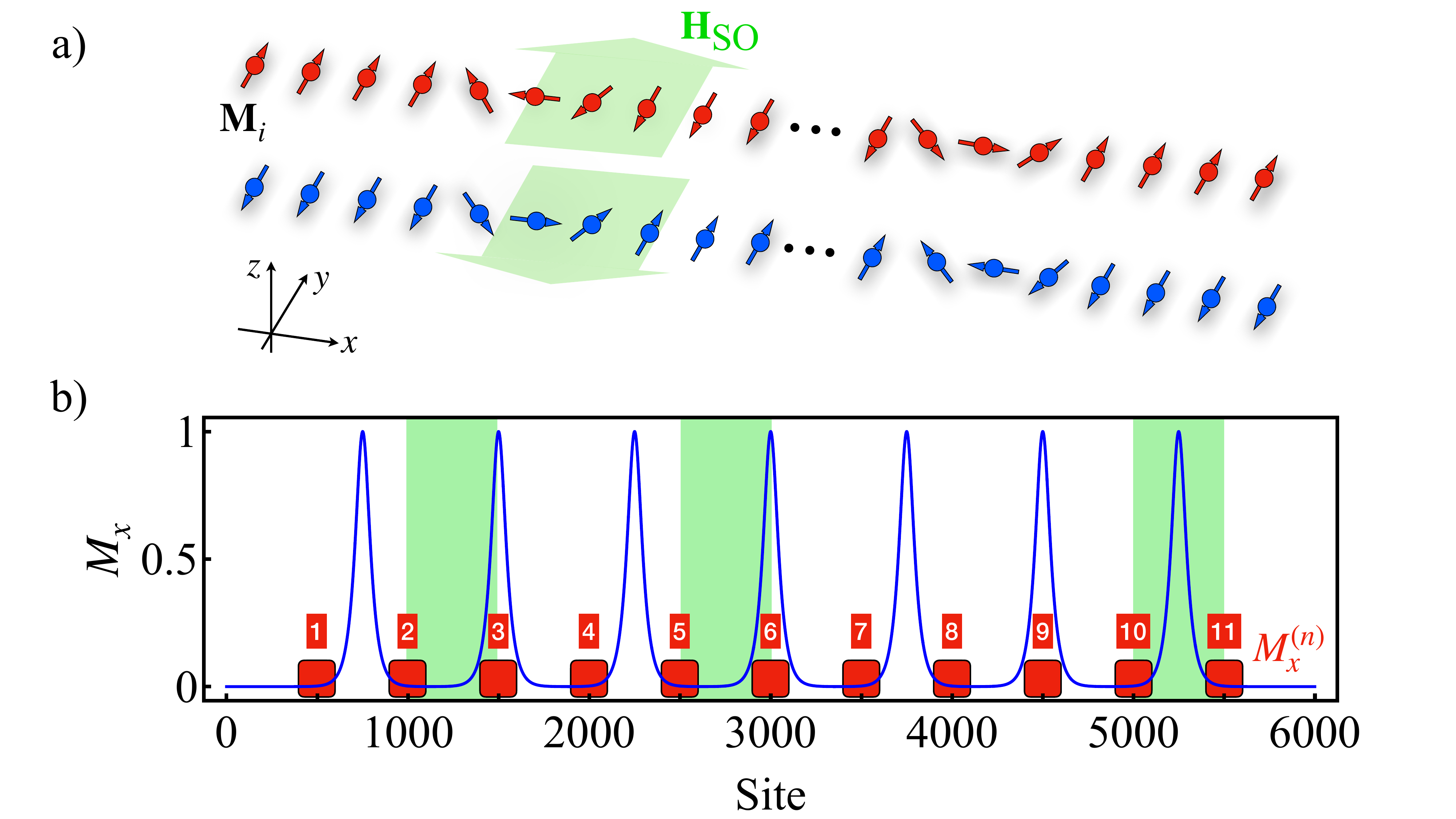}
\caption
{
}
\label{fig_4}
\end{figure}

\newpage

\begin{figure}[h!]
\hspace*{-0cm}\includegraphics[width=.9\textwidth]{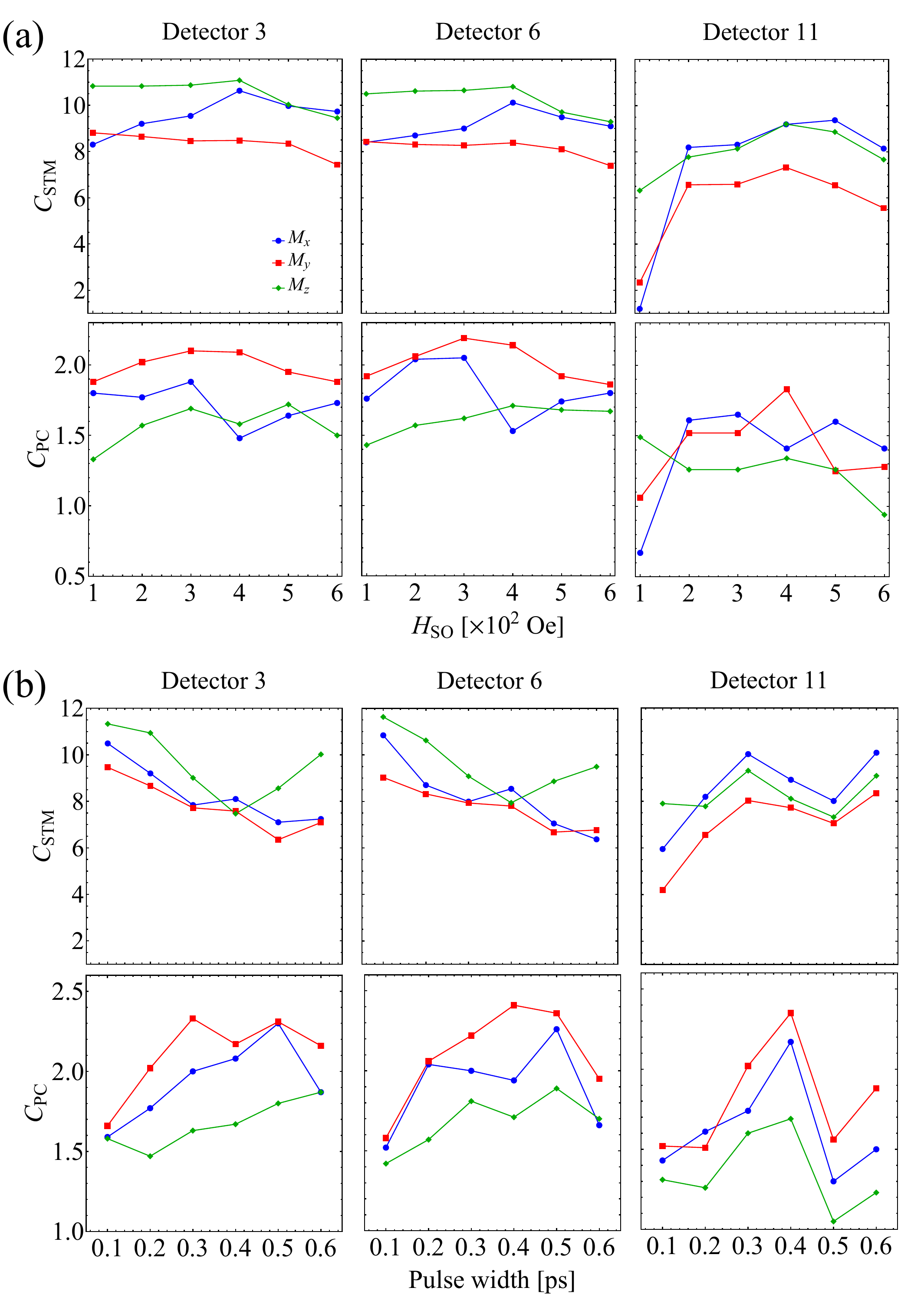}
\caption
{
}
\label{fig_5}
\end{figure}


\begin{figure}[h!]
\hspace*{-0cm}\renewcommand{\figurename}{}\setcounter{figure}{3}  
\caption
{
}
\label{supp_fig_4}
\end{figure}

\end{document}


\title{Supplementary Material: Chaotic Proliferation of Relativistic Domain Walls for Reservoir Computing}

\author{J. A. V\'elez}
\email{jvelez008@ikasle.ehu.eus}
\affiliation{Donostia International Physics Center, 20018 San Sebasti\'an, Spain}
\affiliation{Polymers and Advanced Materials Department: Physics, Chemistry, and Technology, University of the Basque Country, UPV/EHU, 20018 San Sebasti\'an, Spain}

\author{M.-K. Lee}
\affiliation{Department of Applied Physics, Waseda University, Okubo, Shinjuku-ku, Tokyo 169-8555, Japan.}

\author{G. Tatara}
\affiliation{RIKEN Center for Emergent Matter Science (CEMS) and RIKEN Cluster for Pioneering Research (CPR), 2-1 Hirosawa, Wako, Saitama, 351-0198 Japan}

\author{P.-I. Gavriloaea}
\affiliation{Instituto de Ciencia de Materiales de Madrid, CSIC, Cantoblanco, 28049 Madrid, Spain\looseness=-1}

\author{J. Ross}
\affiliation{School of Physics, Engineering and Technology, University of York, York YO10 5DD, UK.}

\author{D. Laroze}
\affiliation{Instituto de Alta Investigación, CEDENNA, Universidad de Tarapacá, Casilla 7D, Arica, Chile.\looseness=-1}
 
\author{U. Atxitia}
\affiliation{Instituto de Ciencia de Materiales de Madrid, CSIC, Cantoblanco, 28049 Madrid, Spain\looseness=-1}

\author{R. F. L. Evans}
\affiliation{School of Physics, Engineering and Technology, University of York, York YO10 5DD, UK.}

\author{R. W. Chantrell}
\affiliation{School of Physics, Engineering and Technology, University of York, York YO10 5DD, UK.}
\author{M. Mochizuki}
\affiliation{Department of Applied Physics, Waseda University, Okubo, Shinjuku-ku, Tokyo 169-8555, Japan.}

\author{R. M. Otxoa}
\email{ro274@cam.ac.uk}
\affiliation{Hitachi Cambridge Laboratory, J. J. Thomson Avenue, Cambridge CB3 0HE, United Kingdom\looseness=-1}
\affiliation{Donostia International Physics Center, 20018 San Sebasti\'an, Spain}

\date{\today}

\newpage

\bigskip

\maketitle
\tableofcontents


\section*{Introduction}

In this supplementary material, we present additional results that complement the main study on the dynamics of domain walls (DWs) in antiferromagnetic (AF) systems and their application in reservoir computing (RC). The additional results focus on providing a more detailed description of the dynamic indicators used in the main text, such as complexity, entropy, and disequilibrium, as well as showcasing further analyses.

\section*{Dynamic Indicators}

\subsection{Fast Fourier Transform (FFT)}

We employed the FFT to decompose the temporal signals of magnetization into their frequency components when oscillatory behaviors are present. This allows us to identify the dominant frequencies and their variations in response to changes in system parameters, which is crucial for distinguishing between periodic and chaotic dynamic behaviors. The FFT analysis is particularly informative in understanding the transitions among these oscillatory states in our AF system, but it is not utilized in cases of stable fixed-point behaviors where oscillations are absent.

In the specific cases we have worked on,  Fig. \textcolor{blue}{2}b shows the Fast Fourier Transform (FFT) of the time series of the $M_x$ component for each magnetic moment along the track, while  Fig. \textcolor{blue}{2}d presents the FFT of the spatial average $\langle M_x \rangle$ along the same track. The calculation of these transforms was carried out using sufficiently long time series to obtain a representative sample, taking into account the limitation of edge effects. As domain wall (DW) proliferation occurs, the dynamic front expands toward the track boundaries, which can distort the measurement when it approaches the edges. To mitigate this effect, a track of 90,000 unit cells was used, and a transient time of 200 ps was established to allow the system to reach a stable dynamic state before measurement, both in chaotic and periodic states.

In the case of chaotic states with DW proliferation, segments of the track were selected where the dynamic front had not yet reached the boundary, resulting in a lower average value of $M_x$. For periodic states, samples were taken after the transient time, once the system had reached a stable periodic behavior.

\subsection{Complexity Indicator}

To determine whether the dynamic state is chaotic or periodic, we use the complexity indicator $C$, expressed as the product of disequilibrium $D$ and entropy $H$. In this study, complexity is calculated from the time series of the component $m_\nu$, where $m_\nu$ represents the average of $M_\nu$ over each component $\nu = x, y$ along the track. Complexity is calculated after allowing a transient time of 200 ps and over a total duration of 1000 ps. This choice is due to the size of the simulated system, which consists of an Mg$_2$Au structure with 90,000 atomic cells along the $x$-axis. The simulations were performed using Fortran, and this duration was chosen to ensure that the system reached a stable final state both in space and time, as well as to provide an adequate sample for numerical analysis. The entropy $H$ measures the uncertainty or disorder in the system and is calculated using the formula:

\begin{equation}
H = -k \sum_{i=1}^{N_p} p_i \log p_i
\end{equation}

where $p_i$ represents the probability of state $i$ and $N_p$ is the total number of possible states. In this context, the constant $k$ does not represent the Boltzmann constant. Instead, it is typically set to $k = 1$ to simplify the calculation, so that entropy is expressed in ``nats" if the natural logarithm is used or in ``bits" if the base-2 logarithm is used. This choice allows entropy to be adapted to information theory without introducing physical units.

In Supplementary Fig.~\ref{Supp_fig_1}, this probability $p_i$ is derived from the normalized time series of $|m_x(t)-\langle m_x\rangle|$.  The calculation of microstates $p_i$ follows a specific process: First, the time series of the average $m_{x} (t) = \langle M_x (t) \rangle$ is taken and translated by subtracting the mean value $\langle m_x \rangle$ from each point in the series, yielding a time series centered at zero. Subsequently, the absolute value of each point is calculated, ensuring that all values are positive and reducing information loss associated with minima. The resulting time series is normalized to the range [0, 1] to facilitate the identification of binary microstates.

Peaks in the normalized time series are identified (marked by red dots in the figure). To classify each peak as 0 or 1, a threshold corresponding to the mean value of all detected peaks is used, represented by the green line in the figure. A binary value of 0 is assigned if the peak is below this mean, and 1 if it is above. The resulting binary string, of length $n_{\text{bin}}$, represents a specific microstate and is converted to a base-10 number to construct the histogram of microstates, as shown in the figure panels.

The disequilibrium $D$ measures the deviation of the probability distribution from a uniform equilibrium and is calculated as:

\begin{equation}
D = \sum_{i=1}^{N_p} \left(p_i - \frac{1}{N}\right)^2
\end{equation}

where $N$ is the number of possible states, which in this case corresponds to the number of bins in the histogram. Disequilibrium $D$ reflects how different the distribution of $p_i$ is compared to a uniform distribution, where each state would have the same probability $1/N$. Finally, complexity $C$ is defined as the product of entropy $H$ and disequilibrium $D$ \cite{Perez2023,Lopez1995}:

\begin{equation}
C = H \cdot D.
\end{equation}

\begin{figure}[h]
\centering
\includegraphics[width=15cm]{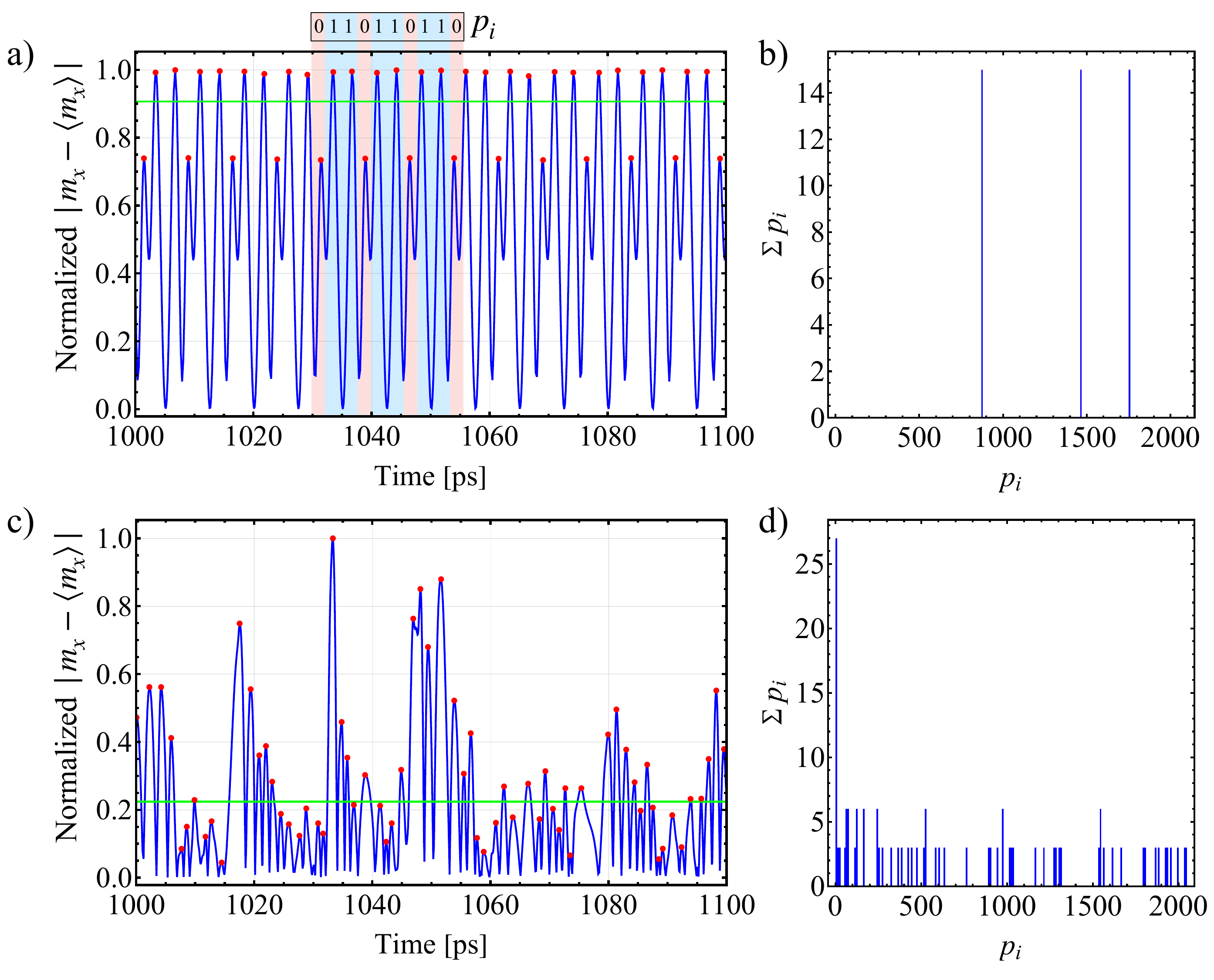}\renewcommand{\figurename}{Supplementary Fig.}\setcounter{figure}{0}
\caption{\textbf{Time Series and Microstate Histograms for Periodic and Chaotic States.} Panels (a) and (b) show the time series and histogram of microstates for a periodic state, with $H_{\text{so}} = 750$ Oe and an oscillation period of $T = 15$ ps. In this periodic regime, the time series exhibits a regular pattern, and the microstate histogram shows an equal distribution among microstates. Panels (c) and (d) display the time series and histogram of microstates for a chaotic state, with $H_{\text{so}} = 780$ Oe and an oscillation period of $T = 25$ ps. The chaotic regime results in an irregular time series, and the histogram indicates a wider, uneven distribution of microstates.}
\label{Supp_fig_1}
\end{figure}

To calculate the total complexity, we consider the normalized time series of each component $|m_x(t)-\langle m_x\rangle|$ for $\nu = x, y$. Based on this, we calculate the complexity associated with each component and then take the average to obtain the total complexity:

\begin{equation}
C_{\text{total}} = \frac{C_x + C_y}{2}.
\end{equation}

The component $M_z$ is not considered in this analysis. Supplementary Fig.~\ref{Supp_fig_1}a) illustrates the processing of each component, in this case $M_x$, showing the steps taken to obtain the data required for the calculation of complexity $C_x$. The mean value is used as a threshold for classifying peaks, which ultimately contributes to the calculation of $C$ for each component. The average of these complexity values provides a comprehensive measure of dynamics of the system. A value of $C > 0$ indicates a chaotic state (represented by blue points in Fig. 3c in the main text), while $C = 0$ indicates a periodic state (red points in Fig. 3c).

To further illustrate the calculation of complexity, Supplementary Fig.~\ref{Supp_fig_1} shows the time series and histograms of microstates for both periodic and chaotic states. For the periodic state, with $H_{\text{so}} = 750$ Oe and an oscillation period of $T = 15$ ps, the calculated values are an entropy $H_x = 1.09861$, a disequilibrium of $D_x = 0$, and consequently, a complexity of $C_x=0$, as expected for a system with regular, predictable behavior. In contrast, for the chaotic state, with $H_{\text{so}} = 780$ Oe and an oscillation period of $T = 25$ ps, the complexity analysis yields an entropy of $H_x = 3.76561$, a disequilibrium of $D_x = 0.0145063$, and a resulting complexity of $C_x = 0.0546248$. These values confirm the increased disorder and deviation from uniformity in the chaotic state, as reflected by the higher entropy and non-zero complexity.

\subsection{Bifurcation diagrams}

Bifurcation diagrams are graphical tools used to study changes in the qualitative behavior of a system as a control parameter is varied. In this study, the control parameters are the amplitude of the spin-orbit field $H_{\text{SO}}$ and the excitation period $T$. By varying these parameters, we analyze the transitions between different dynamic behaviors of the DWs. The diagrams as shown in Fig.~3 in the main text illustrate the state of the system, showing periodic or chaotic behavior as indicated by the complexity and the growth rate of new DWs, $\xi$. This approach helps to visualize how changes in $H_{\text{SO}}$ and $T$ influence the DW dynamics, revealing critical insights into the mechanisms driving the behavior of the system and aiding in the identification of optimal operating regimes for spintronic and neuromorphic applications.

\section{Additional Results and Complementary Analysis}

In this section, we present additional results that complement the main study. Supplementary Fig.~\ref{Supp_fig_2} illustrates the dynamic states of the system for four different excitation periods ($T = 15, 20, 25, 30 \, \text{ps}$), highlighting both periodic and chaotic behaviors under various conditions of the spin-orbit field ($H_{\text{SO}}$).

\begin{figure*}[h]
	\centering
	\includegraphics[width=\textwidth]{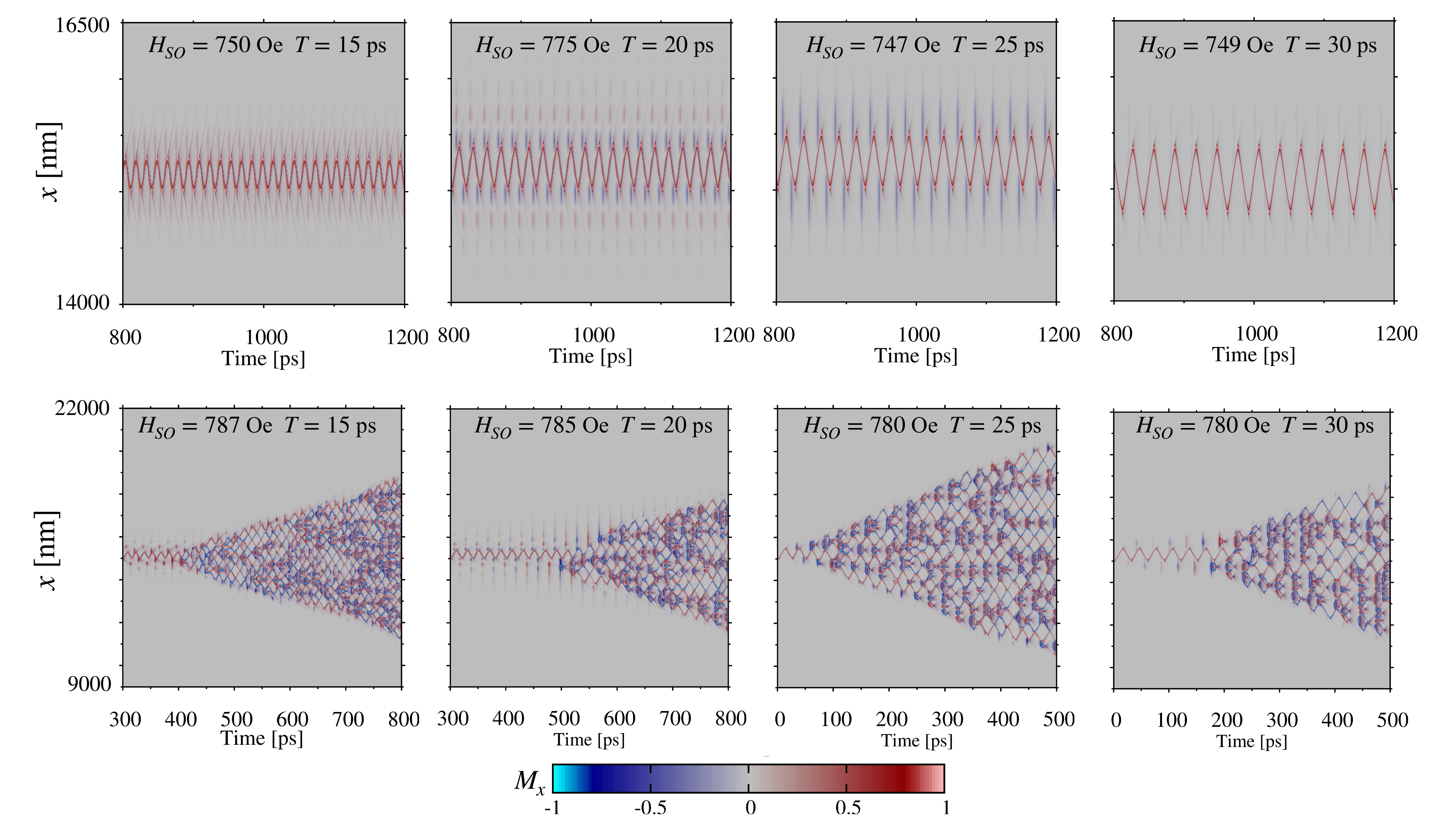}\renewcommand{\figurename}{Supplementary Fig.}\setcounter{figure}{1}
	\caption{Dynamic states of the system as a function of excitation period $T$. The top row illustrates periodic states, while the bottom row displays chaotic states. Each column corresponds to a different excitation period, increasing from left to right. The transition from periodic to chaotic behavior is observed as $T$ and $H_{\text{SO}}$ vary, demonstrating the sensitivity of the system’s dynamics to these parameters.}
	\label{Supp_fig_2}
\end{figure*}

Each subplot represents the $M_x$ component dynamics of the DW along the $x$-axis. In the periodic states (top row), the DW exhibits regular oscillations with minimal emission of spin waves. These states show stability with well-defined periodic behavior, indicating an orderly dynamic regime. In contrast, the chaotic states (bottom row) display a complex interplay of DW nucleation and proliferation, significantly influenced by the higher spin-orbit field values. The high emission of spin waves for lower $T$ values is evident, contributing to the chaotic dynamics. Notably, the number of newly formed DWs increases with higher $H_{\text{SO}}$, reflecting the dynamic instability of the system. 

Further analysis of the chaotic states reveals specific trends associated with different values of $T$: For $T = 15~\text{ps}$, the chaotic regime exhibits rapid nucleation and a high number of DWs, indicating a highly unstable state. At $T = 20~\text{ps}$, a similar level of DW activity is observed, though slightly less intense than at $T = 15~\text{ps}$. As $T$ increases to $25~\text{ps}$, the system continues to show chaotic behavior, but the rate of DW proliferation is somewhat reduced compared to the states at $T = 15$ and $20~\text{ps}$. Finally, at $T = 30~\text{ps}$, the chaotic state demonstrates a mixture of high and low DW activity regions, suggesting intermittent chaos.

It is worth noting that in the bottom panels of Supp. Fig.~\ref{Supp_fig_2}, both the amplitude of $H_{\text{SO}}$ and the period $T$ vary, making it difficult to isolate the influence of a single factor. However, an in-depth analysis of the individual effects of $H_{\text{SO}}$ and $T$ is presented in the main text, where each parameter is varied independently. This complementary study provides further insight into how each parameter contributes uniquely to the transition between periodic and chaotic states, confirming that lower $T$ values and higher $H_{\text{SO}}$ amplitudes together drive the system towards a regime of rapid DW proliferation.

These observations underscore the significant role of the excitation period and spin-orbit field in dictating the dynamic behavior of DWs in antiferromagnetic systems. The transition from periodic to chaotic states is marked by an increase in spin wave emissions and DW nucleation, leading to complex and highly nonlinear dynamics.
To determine whether the dynamic state is chaotic or periodic, we use the complexity indicator, which is defined as the product of disequilibrium and entropy. Supplementary Fig. \ref{Supp_fig_3} illustrates the relationship between the excitation period $T$ and the dynamic state of the system under a fixed spin-orbit field $H_{\text{SO}} = 750$ Oe, highlighting how adjusting the excitation period can control dynamic states, alternating between chaotic and periodic states.

\begin{figure}[h]
    \centering
    \includegraphics[width=\textwidth]{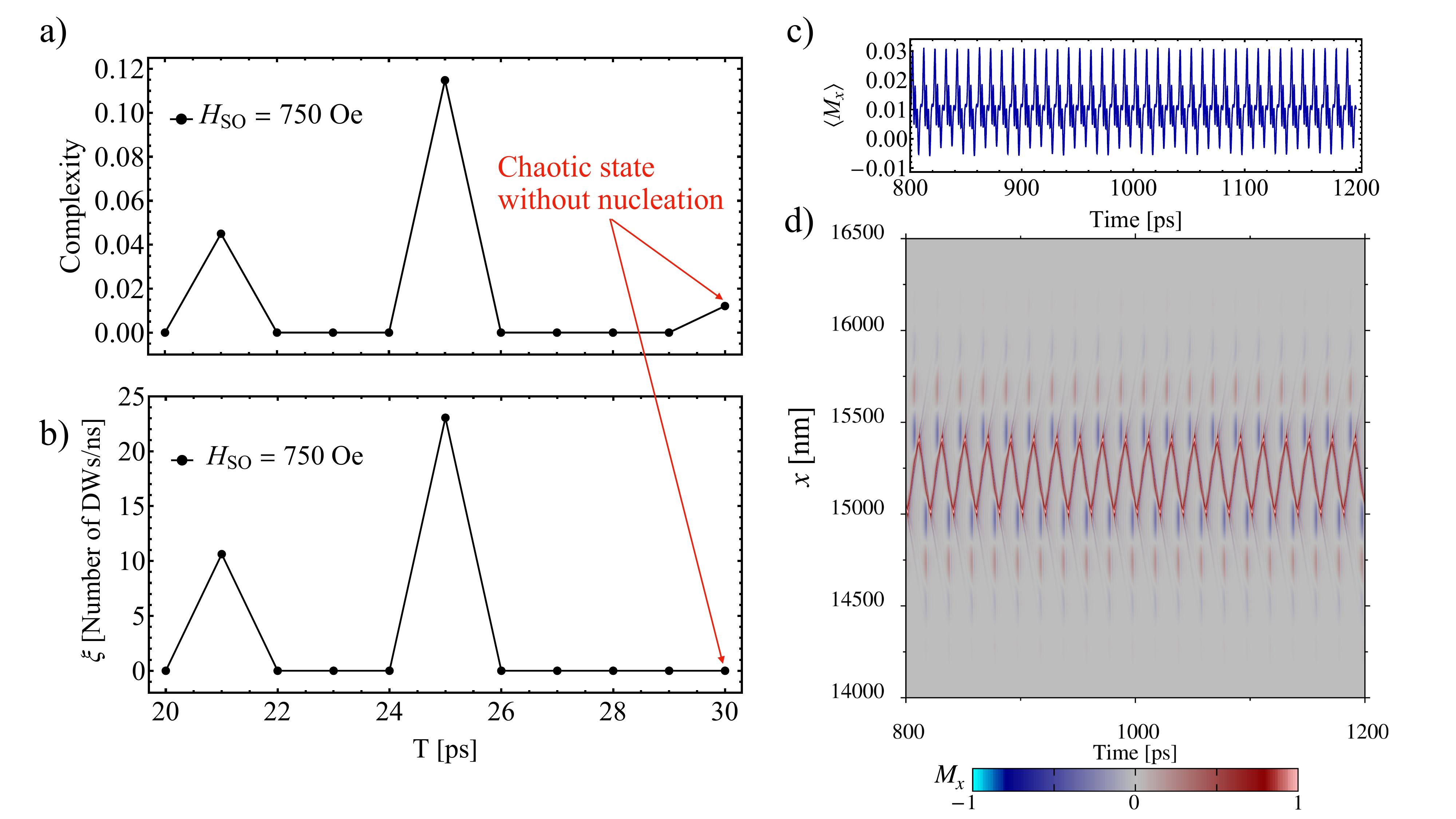}\renewcommand{\figurename}{Supplementary Fig.}\setcounter{figure}{2}
    \caption{Analysis of complexity and $\xi$ as a function of the excitation period $T$ for a fixed spin-orbit field $H_{\text{SO}} = 750$ Oe. In panel (a), the system complexity $C$ is shown as a function of the excitation period $T$ with a field of $H_{\text{SO}} = 750$ Oe. Moving to panel (b), the growth rate of new DWs $\xi$ is presented as a function of the excitation period $T$ for the same field. Both panels highlight the point at $T = 30$ ps, where a state with positive complexity but no DW nucleation is observed, indicating a chaotic state without DW proliferation. Next, in panel (c), the time series of the average value of $M_x$ is shown for $T = 15$ ps and $H_{\text{SO}} = 786$ Oe, corresponding to a chaotic state without nucleation. Finally, panel (d) illustrates the dynamics of the $M_x$ component for each magnetic moment of the system along the $x$ axis for a section of the track, also for $T = 15$ ps and $H_{\text{SO}} = 786$ Oe. This state is particularly useful for observing the effect of chaos without nucleation, as the complexity algorithm is very sensitive and can detect these states.}
    \label{Supp_fig_3}
\end{figure}

In panel (\ref{Supp_fig_3}a), we observe the calculated complexity as a function of $T$. Positive complexity values (blue dots in figure 3d in the main text) indicate chaotic states, while zero complexity values (red dots in figure 3d in the main text) indicate periodic states. This analysis shows that the system can transition between chaos and periodicity by modifying the excitation frequency. Panel (\ref{Supp_fig_3}b) displays the growth rate of new DWs, $\xi$, as a function of $T$. Here, $\xi = 0$ signifies no nucleation, while $\xi > 0$ reflects DW nucleation, indicative of chaos induced by relativistic effects. Notably, at $T = 30$ ps, we see a state with $\xi = 0$ but positive complexity, suggesting a chaotic state without DW nucleation.

To further illustrate this unique case, panels (\ref{Supp_fig_3}c) and (\ref{Supp_fig_3}d) show the dynamics of $M_x$ in a chaotic state without nucleation. Although these panels correspond to $T = 15$ ps and $H_{\text{SO}} = 786$ Oe, they were selected to clearly show the complex dynamics without DW proliferation. This is because the complexity algorithm used is highly sensitive and capable of detecting these states. The choice of these specific parameters allows for detailed visualization of the chaotic behavior, showing how positive complexity can exist without an increase in the number of DWs.

Panel (\ref{Supp_fig_3}c) presents the average magnetization $\langle M_x \rangle$ over time for $T = 15$ ps and $H_{\text{SO}} = 786$ Oe, showing the global dynamic behavior of the system. Panel (\ref{Supp_fig_3}d) provides a detailed view of the $M_x$ component dynamics for each magnetic moment along a section of the track at $T = 15$ ps. This panel reveals intricate spin interactions and localized dynamic behaviors, emphasizing the complexity and richness of the system response under specific conditions.

\section{Reservoir Computing}

\begin{figure*}[tb]
	\centering
	\includegraphics[scale=0.55]{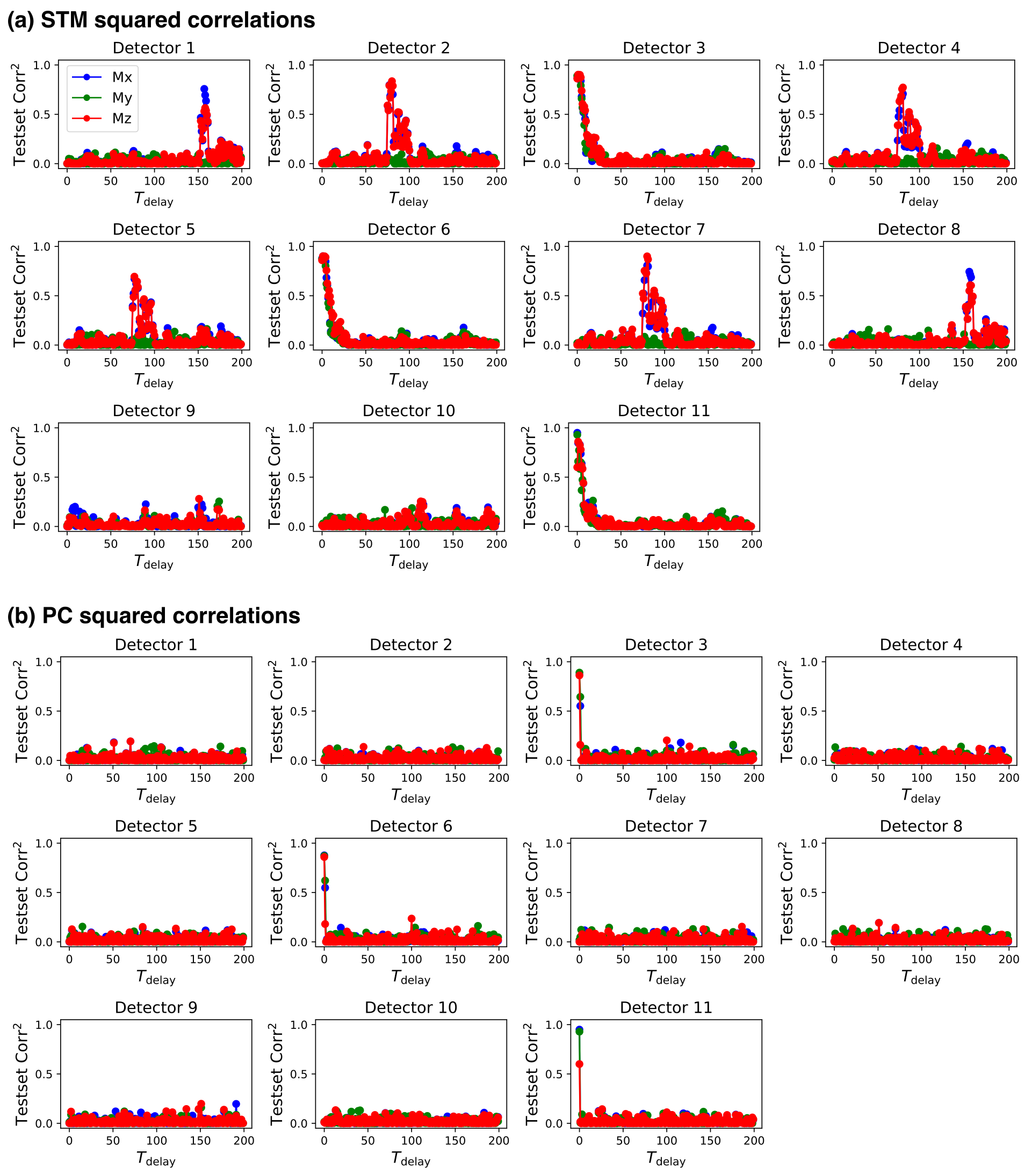}\renewcommand{\figurename}{Supplementary Fig.}\setcounter{figure}{3}
	\caption{Squared correlations (Corr$^2$) for the testing dataset carried out by each of the eleven detectors for (a) STM and (b) PC tasks.}
	\label{Supp_fig_4}
\end{figure*}
As outlined in the main text, we consider two one-dimensional ferromagnetic layers with mutual AF exchange coupling, as illustrated in Fig.~4a in the main text. Each neighboring DWs have opposite winding numbers to mimic the texture generated by the chaotic proliferation from a single seed DW which preserves the topological number. For simplicity, we only consider the current-induced staggered spin-orbit field, with the spin transfer torque being ignored, and the $J_1$ AF exchange coupling in Mn$_2$Au as defined in \cite{Otxoa2020b} has been omitted.

The squared correlation for each detector with $H_{\text{SO}}=20$~mT, pulse width $p=0.2$~ps, Gilbert damping $\alpha=0.001$, and number of virtual nodes $N_{\rm vn}=100$, is shown in Supplementary Fig.~\ref{Supp_fig_4} with different colors denoting the results by using different components $j=x,y,z$ of the magnetization to form the reservoir state vector $\mathbf{R}^{(n)}_j(T_i)$. 
The results vary significantly depending on the detector location. Empirical testing with different input area locations for the two tasks reveals that when a detector is placed close to both an edge of any input areas and to one of the DW centers [e.g., detectors 3 and 6 in Fig.~4b in the main text] or near the system edge [e.g., detectors 11 in Fig.~4b in the main text], its squared correlations for both STM and PC tasks are close to 1 when $T_{\rm delay}=0$, and then decay monotonically as $T_{\rm delay}$ increases. This pattern indicates the presence of fading memory in STM task, which is essential for physical reservoirs to perform repeatable computations, since it guarantees that a sufficiently long input sequence will erase the previous memory carried by the reservoir~\cite{Jaeger2001}. On the other hand, for detectors not positioned close to the above mentioned areas, the squared correlation is already much less than 1 for $T_{\rm delay}=0$. This indicates that these detectors cannot be trained to predict the desired target functions. The requirements for the detector positioning highlights the need for both a large spatial gradient of $H_{\rm SO}$ (e.g., locations near the edges of input local field areas) and a large gradient of $M_y$ (e.g., near DW center) or strong spin wave reflection near the system edge. These factors are crucial for exciting significant magnetization responses that possess memory and nonlinearity relative to the input.

For detectors situated far from the edges of input areas, DW centers, or the system edge, no obvious correlations are found in PC task for all the delay times investigated from $T_{\rm delay}=0$ to $200$. Interestingly, in the STM task, these detectors exhibit a peak in squared correlations at a finite delay time. For instance, for detector-7 and detector-8 in Supplementary Fig.~\ref{Supp_fig_4}a), their STM squared correlations show peaks located at $T_{\rm delay}\approx 80$ and $150$, respectively. For delay times beyond these peak instants, the correlations again decay to nearly zero. 
This behavior is attributed to the spin wave propagation. The local spin-orbit field excites magnetization responses near its applied area, then the spin waves propagate to positions of farther detectors, resulting in a delayed memory carried by these detectors. 
This behavior may find applications that require the reservoir to reconstruct past inputs that were injected to the system a long time ago. The peak instants of these delayed memory can naturally be tuned by changing the location of detectors. To confirm this spin wave scenario, we have carried out the tasks using a larger Gilbert damping constant $\alpha=0.5$ and found that the peaks of STM correlations at finite $T_{\rm delay}$ disappeared. This result aligns with our expectations, as the spin waves already decay to small amplitudes before reaching detectors located farther from the input areas when the Gilbert damping is large.

We note that, after injecting an input sequence of over 1000 random digits into the DW array, the original positions of DWs slightly shift. Following a sufficiently long relaxation period of more than 200 ps, these shifted DWs cannot return back to their original positions, as our model lacks pinning sites or magnetic impurities that tend to drag or attract the DWs. However, when using this newly relaxed configuration and applying the same input series once again, the final DW positions become nearly fixed. Importantly, the resultant STM and PC capacities remain comparable with the first-round results obtained from the initial DW configuration. Thus, within the designed input procedure in our work, the repeatability of this DW system is fulfilled. 
Moreover, we compared the performance by this DW array with that of a purely AF state without DW textures —referred to as the ``non-reservoir" state, where $M_x=M_z=0, M_y=-1 (+1)$ for all sites in the upper (lower) layers. In this non-reservoir state, the random sequence of spin-orbit field input fails to generate significant magnetization responses due to the initially saturated $M_y$ configuration. As a result, this state lacks necessary functionalities for short-term memory and nonlinearity. Consequently, the DW array demonstrates significantly better performance compared to the purely AF state without magnetic textures.

\section{Conclusion}

In conclusion, this supplementary document has provided a deeper insight into the dynamics of domain walls (DWs) in antiferromagnetic systems and their potential application in reservoir computing (RC). Through the implementation of dynamic indicators such as complexity, entropy, and disequilibrium, we have been able to clearly differentiate between chaotic and periodic states, offering a detailed understanding of the underlying mechanisms governing these dynamic transitions.

We have demonstrated how adjusting parameters such as the excitation period $T$ and the spin-orbit field $H_{\text{SO}}$ can control the transition between periodic and chaotic states, highlighting the sensitivity of the system to these parameters and its potential for applications in spintronics and neuromorphic computing.

Additionally, we have presented specific results showing the high emission of spin waves and the increase in DW proliferation under certain conditions, emphasizing the complexity and richness of the system behavior. These findings not only expand our fundamental understanding of DW dynamics in antiferromagnetic systems but also open new avenues for exploring advanced applications in emerging technologies.

\bibliographystyle{naturemag}
\bibliography{bib2}
\vspace*{1cm}